\documentclass[12pt]{article}

\usepackage{amsmath}
\usepackage{graphicx}
\usepackage{multirow}
\usepackage{epsfig}
\usepackage{rotating}

\newcommand{\ba}{\begin{array}}
\newcommand{\ea}{\end{array}}

\begin{document}

\begin{flushright}
PITHA 09/26\\
\today
\end{flushright}
\vspace*{0.8truecm}

\begin{center}
{\LARGE \bf \boldmath Constraints on the mass and mixing of $Z'$ bosons}

\vspace*{0.8truecm}
{\large
J. Erler$^{a,b}$\footnote{erler@fisica.unam.mx},
P. Langacker$^c$\footnote{pgl@ias.edu},
S. Munir$^a$\footnote{smunir@fisica.unam.mx},
E. Rojas$^a$\footnote{eduardor@fisica.unam.mx}}

\vspace*{0.5truecm}
{\footnotesize \it
$^a$ Departamento de F\'isica Te\'orica, Instituto de F\'isica, Universidad Nacional\\ Aut\'onoma de M\'exico, 04510 M\'exico D.F., M\'exico\\
$^b$ Institut f\"ur Theoretische Physik E, RWTH Aachen,
52056 Aachen, Germany\\
$^c$ School of Natural Sciences, Institute for Advanced Study, Einstein Drive,\\ Princeton, NJ 08540, USA}
\end{center}

\begin{abstract}
\footnotesize
\noindent We tested several models in which the Standard Model (SM) gauge group
is extended by an additional $U(1)$ gauge symmetry against available
electroweak precision data to impose limits on the mass of the neutral
$Z'$ boson, $M_{Z'}$, predicted in all such models, and on the $Z-Z'$
mixing angle, $\theta_{ZZ'}$,  at 95\% C.L. We found lower limits on
$M_{Z'}$ of order $1~{\rm TeV}$ in most cases, while $\theta_{ZZ'}$ was found to be constrained to very small values.\\

\textbf{PACS:} 2.60.Cn Extensions of electroweak gauge sector, 14.70.Pw Other gauge bosons

\end{abstract}

\section{Introduction}

One of the best motivated extensions of the SM is a neutral gauge
sector with an extra $U(1)$ symmetry in addition to the SM hypercharge
$U(1)_Y$ and an associated $Z'$ gauge boson, predicted in most Grand
Unified Theories (GUTs) such as $SO(10)$ or
$E_6$~\cite{Georgi:1974sy}. There is an extensive range of models with
such an additional $U(1)$ symmetry (for reviews,
see~\cite{Langacker:2008yv,Rizzo:2006nw,Leike:1998wr} and references therein). Among these,
models based on the $E_6$ GUT group and left-right symmetry groups
have been extensively pursued in the literature and are particularly
significant from the point of view of LHC phenomenology. Here we
summarize the analysis done in~\cite{Erler:2009jh} of various $Z'$ models,
which include $E_6$ based $Z\chi$, $Z_\psi$,
$Z_\eta$ ~\cite{Robinett:1982tq, Langacker:1984dc, Hewett:1988xc, Candelas:1985en}, $Z_N$~\cite{Ma:1995xk,Kang:2004ix}
and $Z_R$~\cite{Erler:1999nx} bosons, a $Z_{LR}$ boson appearing in
models with left-right symmetry (reviewed in Ref.~\cite{Mohapatra:1986uf}),
 a sequential $Z_{SM}$ boson and a family non-universal $Z_{string}$
 boson~\cite{Chaudhuri:1994cd}. 

\section{Mixing and Parameters}

The mixing between the mass eigenstates of $Z'$ and the $Z$ is given by,
\begin{eqnarray}
\tan^2 \theta_{ZZ'} =\frac{M_0^2-M_Z^2}{M_{Z'}^2-M_0^2},
\end{eqnarray}
where $M_0$ is the mass of $Z$ boson in the absence of $Z-Z'$ mixing and is given as 
\begin{eqnarray}
   M_0 = {M_W\over \sqrt{\rho_0} \cos\theta_W}; ~~~~~\rho_0 \equiv \frac{\sum_i (t_i^2 - t_{3i}^2 + t_i) |\langle \phi_i \rangle|^2}
                {\sum_i 2 t_{3i}^2               |\langle \phi_i \rangle|^2}.
\label{wzrelation}
\end{eqnarray}

We set $\rho_0 = 1$ here, which corresponds to a Higgs sector with only $SU(2)$ doublets and singlets. Furthermore, if the $U(1)'$ charge assignments of the Higgs fields, $Q'_i$, are known in a specific model, then there exists an additional constraint~\cite{Langacker:1991pg},
\begin{eqnarray}
  \theta_{ZZ'} = C\ {g_2\over g_1} {M_Z^2\over M_{Z^\prime}^2},
\end{eqnarray}
where $g_1 = g_L/\cos\theta_W$ and where $g_2 = \sqrt{5/3}\ g_1 \sin\theta_W \sqrt{\lambda}$ is the $U(1)^\prime$ gauge coupling. The latter is given in terms of $\lambda$ whose value is conventionally set to 1 here. $C$ is a function of vacuum expectation values (VEVs) of the Higgs fields and the $Q'_i$,
\begin{eqnarray}
  C = - \frac{\sum_i t_{3i} Q^\prime_i |\langle \phi_i \rangle|^2}{\sum_i t_{3i}^2 |\langle \phi_i \rangle|^2}.
\label{paraC}
\end{eqnarray}
For $E_6$ based models one may restrict oneself to the case where the Higgs fields arise from a {\bf 27} representation. The $U(1)'$ quantum numbers are then predicted and eq.~\ref{paraC} receives contributions from the VEVs of three Higgs doublets, $x \equiv \langle\phi_\nu \rangle$, $v \equiv \langle \phi_N \rangle$ and $\bar{v} \equiv \langle \phi_{\bar{N}} \rangle$, respectively, in correspondence with the standard lepton doublet, as well as the two doublets contained in the ${\bf 10}$ of $SO(10) \subset E_6$. They satisfy the sum rule, $|v|^2 + |\bar{v}|^2 + |x|^2 = (\sqrt{2}\ G_F)^{-1} = (246.22 \mbox{ GeV})^2$, and we introduce the ratios,
\begin{eqnarray}
\tau = {|\bar{v}|^2 \over |v|^2 + |\bar{v}|^2 + |x|^2},~~~\omega =
{|x|^2 \over |v|^2 + |\bar{v}|^2 + |x|^2};~~(0 \leq \tau, \omega \leq
1),
\label{Cvalues}
\end{eqnarray}
resulting in different expressions and ranges for $C$ in different models~\cite{Erler:2009jh}.
 
\section{Results}

We used the special purpose FORTRAN package GAPP~\cite{Erler:1999ug} dedicated to the Global Analysis of Particle Properties, wherein the effects of the $Z'$ bosons are taken into account as first order perturbations to the SM expressions. The most stringent indirect constraints on $M_{Z'}$ come from low-energy weak neutral current (WNC) experiments, while the size of the mixing angle $\theta_{ZZ'}$ is strongly constrained by the very high precision $Z$-pole experiments at LEP and SLC. For details on the data, input parameters and analysis, see \cite{Erler:2009jh}.

\begin{table}[t]
\centering
\begin{tabular}{c|r|r|r|r|r|c|c|c} 
\hline
 $Z'$ & \multicolumn{4}{c|}{$M_{Z'}$ [GeV]} & \multicolumn{3}{c|}{$\sin\theta_{ZZ'}$}  & $\chi^2_{\rm min}$ \\ \hline
& EW (this work) & CDF & D\O\ & LEP~2 & $\sin\theta_{ZZ'}$ & $\sin\theta_{ZZ'}^{\rm min}$ & $\sin\theta_{ZZ'}^{\rm max}$ & \\ 
\hline 
$Z_\chi$         & 1,141\phantom{OOO} &    892 & 640 &    673 & $-0.0004$ & $-0.0016$ & 0.0006 & 47.3 \\ \hline
$Z_\psi$         &    147\phantom{OOO} &    878 & 650 &    481 & $-0.0005$ & $-0.0018$ & 0.0009 & 46.5 \\ \hline
$Z_\eta$         &    427\phantom{OOO} &    982 & 680 &    434 & $-0.0015$ & $-0.0047$ & 0.0021 & 47.7 \\ \hline
$Z_N$            &    623\phantom{OOO} &    861 &         &            & $-0.0004$ & $-0.0015$ & 0.0007 & 47.4 \\ \hline
$Z_R$            &    442\phantom{OOO} &            &         &            & $-0.0003$ & $-0.0015$ & 0.0009 & 46.1 \\ \hline
$Z_{LR}$       &    998\phantom{OOO} &    630 &         &    804 & $-0.0004$ & $-0.0013$ & 0.0006 & 47.3 \\ \hline
$Z_{SM}$      & 1,403\phantom{OOO} & 1,030 & 780 & 1,787 & $-0.0008$ & $-0.0026$ & 0.0006 & 47.2 \\ \hline
$Z_{string}$  & 1,362\phantom{OOO} &            &         &            & $ 0.0002$ & $-0.0005$ & 0.0009 & 47.7 \\ \hline
SM                  & \multicolumn{4}{c|}{$\infty$}                               & \multicolumn{3}{c|}{0}                    & 48.5 \\ \hline
\end{tabular}
\caption{95\% C.L. lower mass limits on extra $Z'$ bosons for various
  models from EW precision data and constraints on
  $\sin\theta_{ZZ'}$. For comparison, we show also (where applicable) the limits obtained by CDF, D\O\  and LEP~2.}
\label{Limits}
\end{table}

In Table~\ref{Limits} we present our limits on the $Z'$ parameters for
the models listed in the introduction. Also shown in
Table~\ref{Limits} are the current limits on various $Z'$ boson masses
from the Tevatron and LEP~2. The mass limits at the Tevatron assume
that no decay channels into exotic fermions or superpartners are open
to the $Z'$; otherwise the limits would be moderately weaker. The
Table shows that the mass limits from the EW precision data are
generally competitive with and in many cases stronger than those from
colliders. 

\begin{figure}[h!]
\centering
\begin{tabular}{cc}
\hspace{-20pt}\includegraphics[scale=0.33]{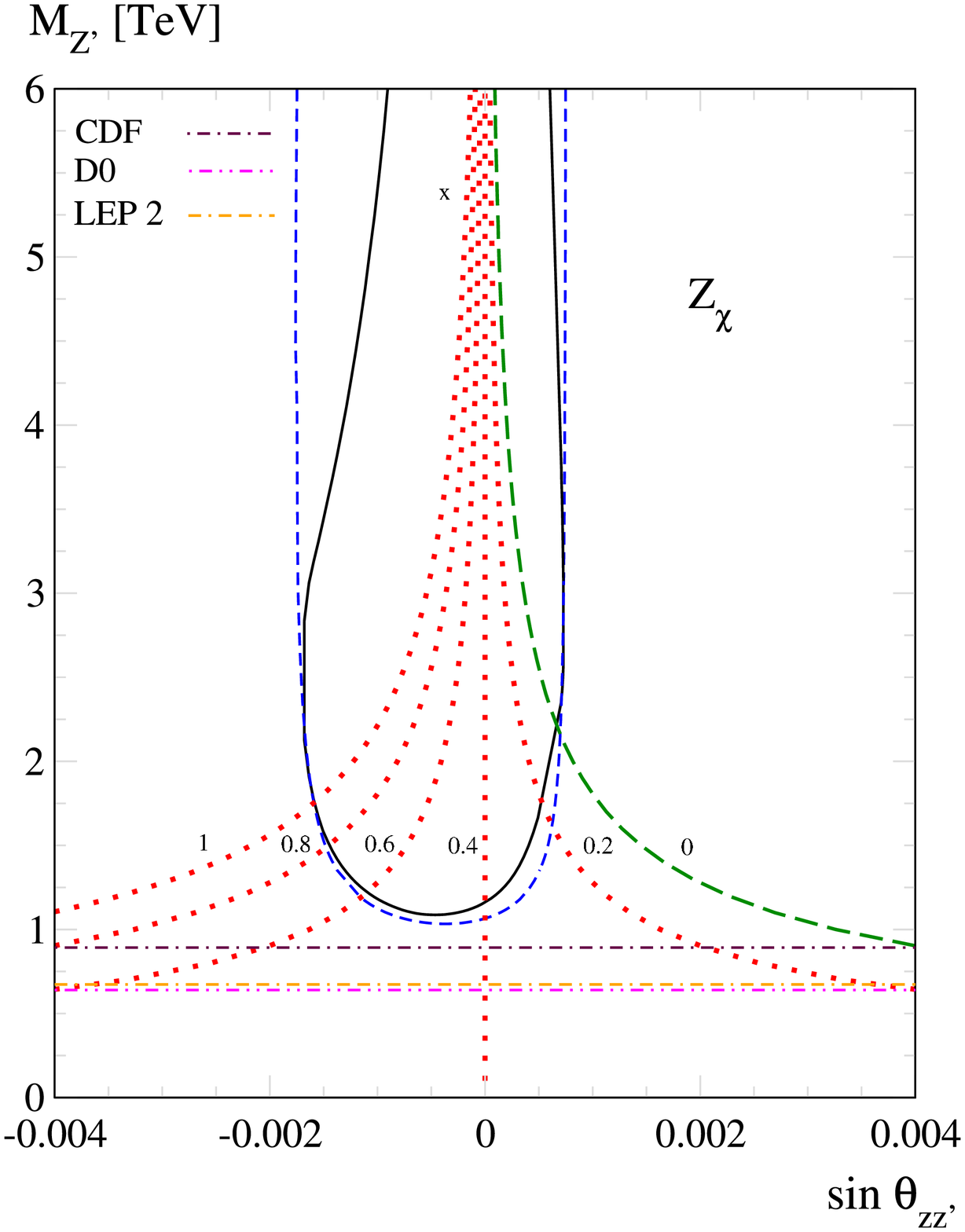} & \hspace{55pt} \includegraphics[scale=0.33]{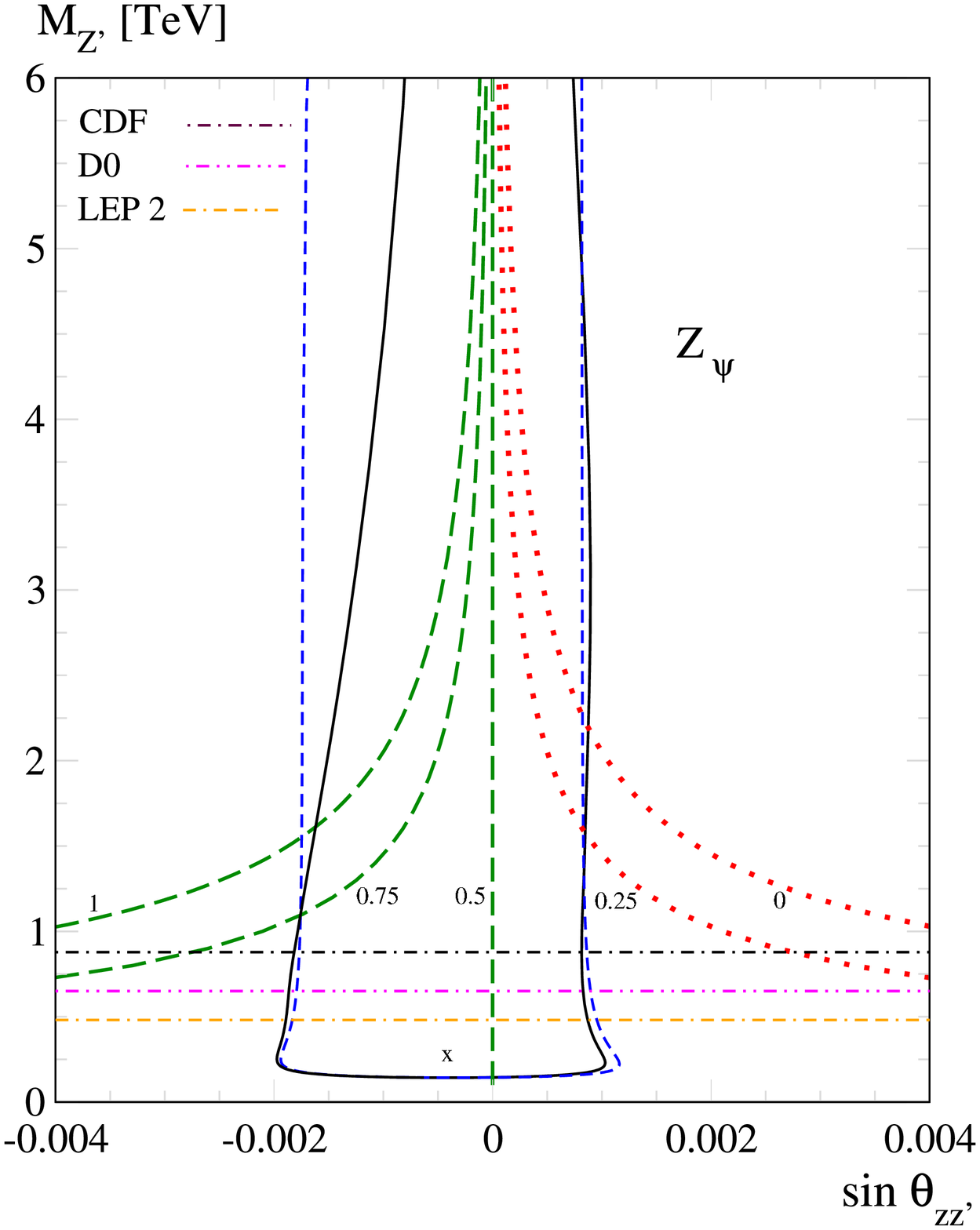} \vspace{90pt} \\
\hspace{-20pt}\includegraphics[scale=0.33]{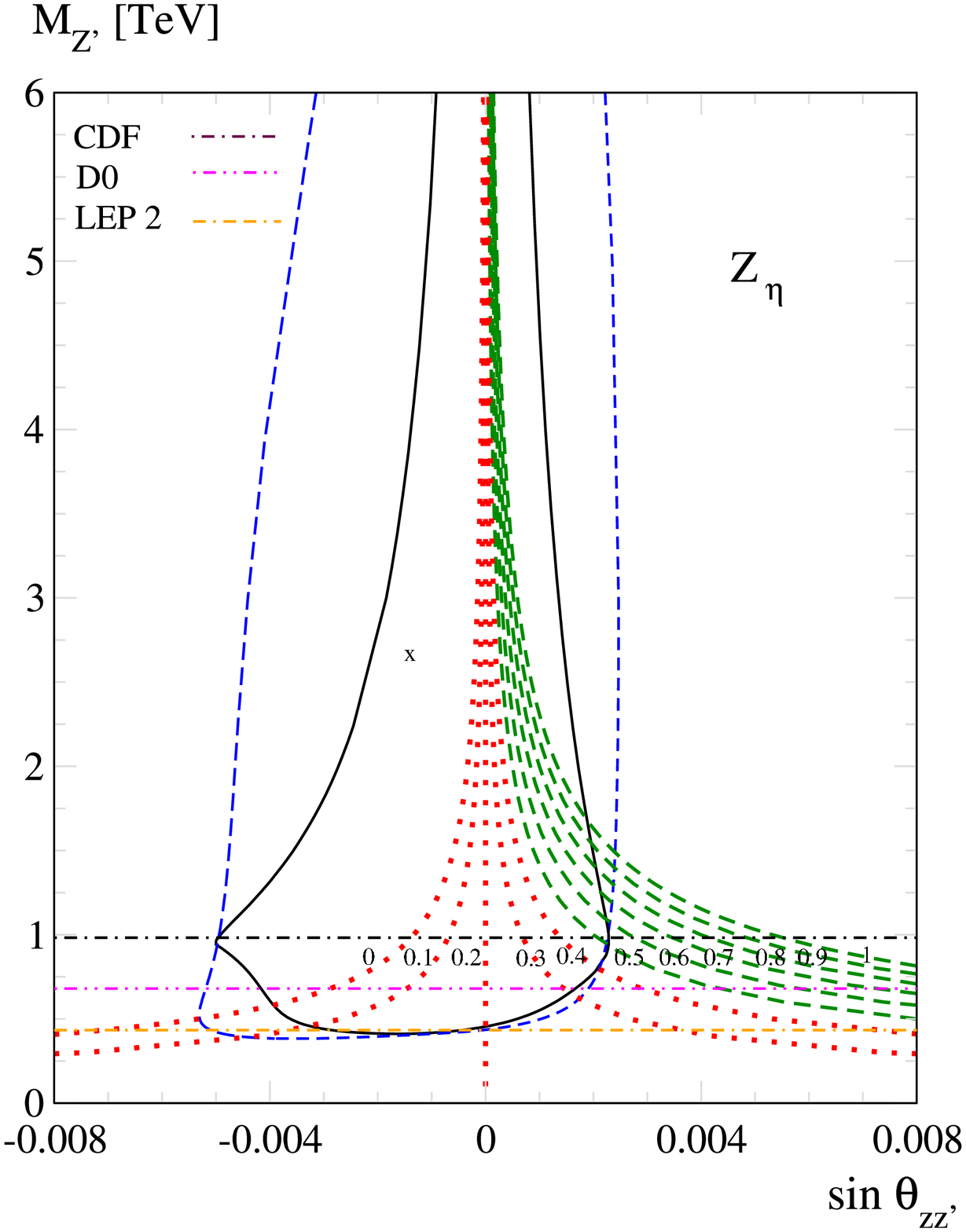} &
\hspace{55pt} \includegraphics[scale=0.33]{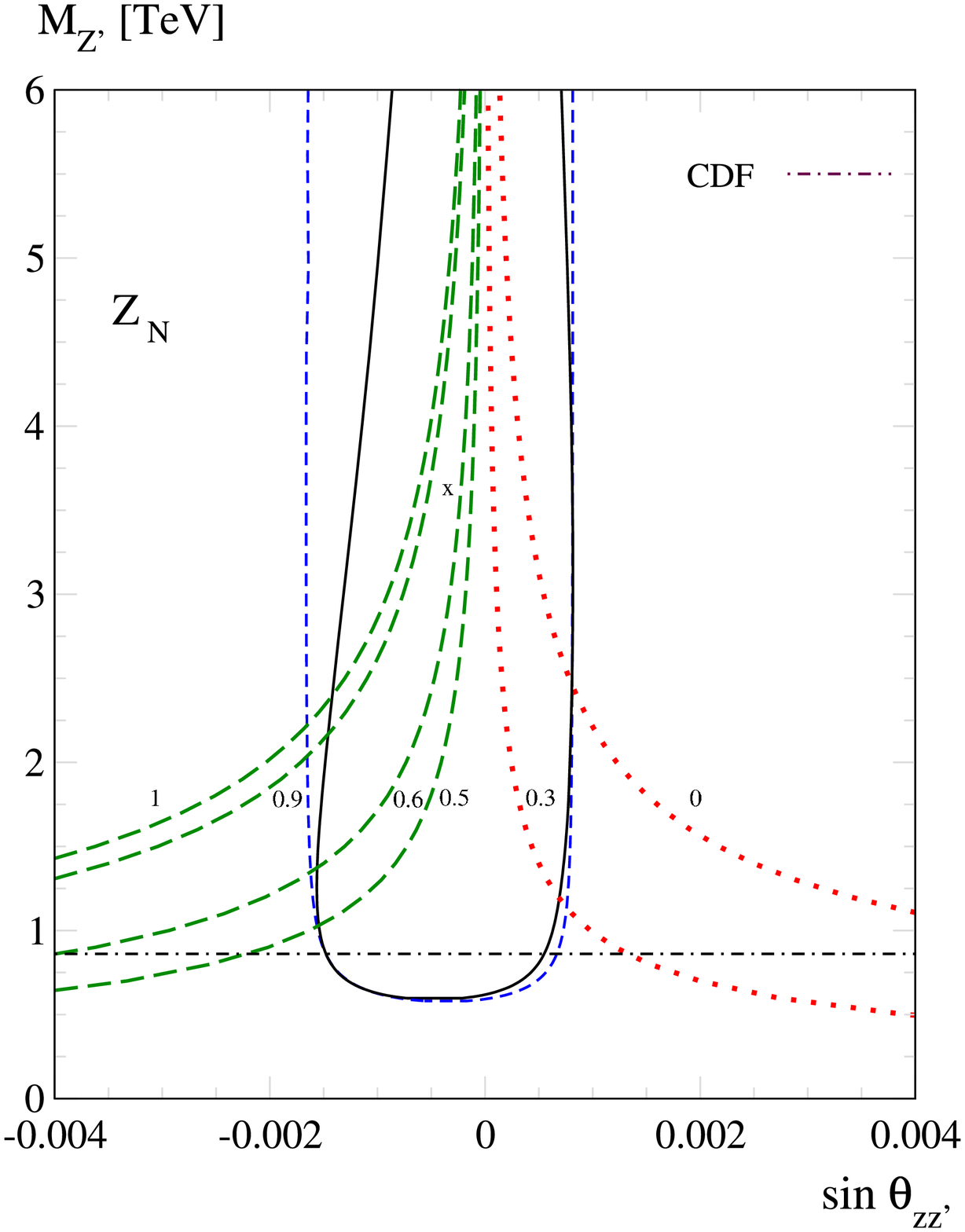} \vspace{10pt} \\
\end{tabular}
\caption{90\% C.L. contours in $M_{Z'}$ vs. $\sin \theta_{ZZ'}$ for
  various models. See the text for details.}
\label{Contours1}
\end{figure}

\begin{figure}[h!]
\centering
\begin{tabular}{cc}
\hspace{-20pt} \includegraphics[scale=0.33]{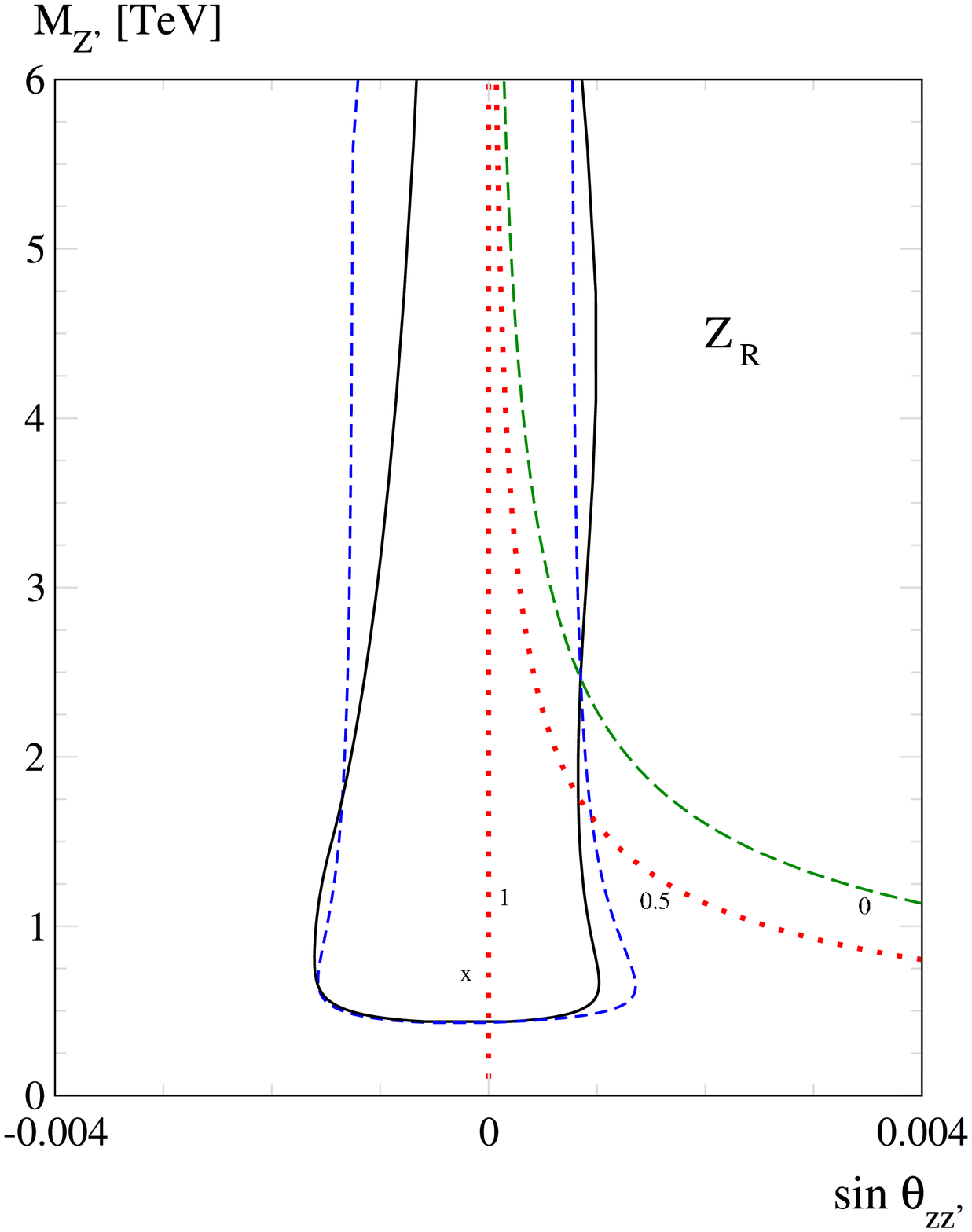} &
\hspace{55pt} \includegraphics[scale=0.33]{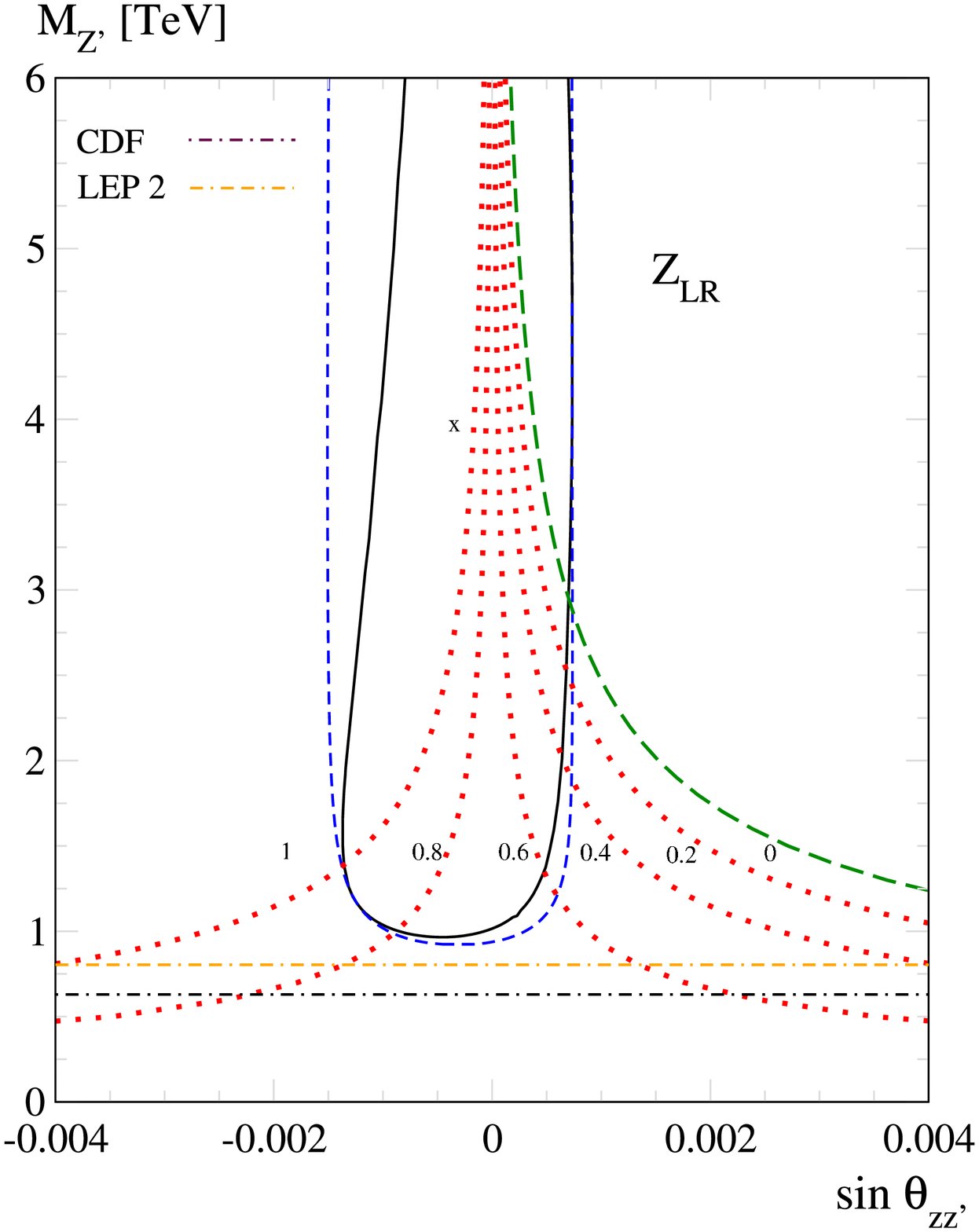} \vspace{90pt} \\
\hspace{-20pt} \includegraphics[scale=0.33]{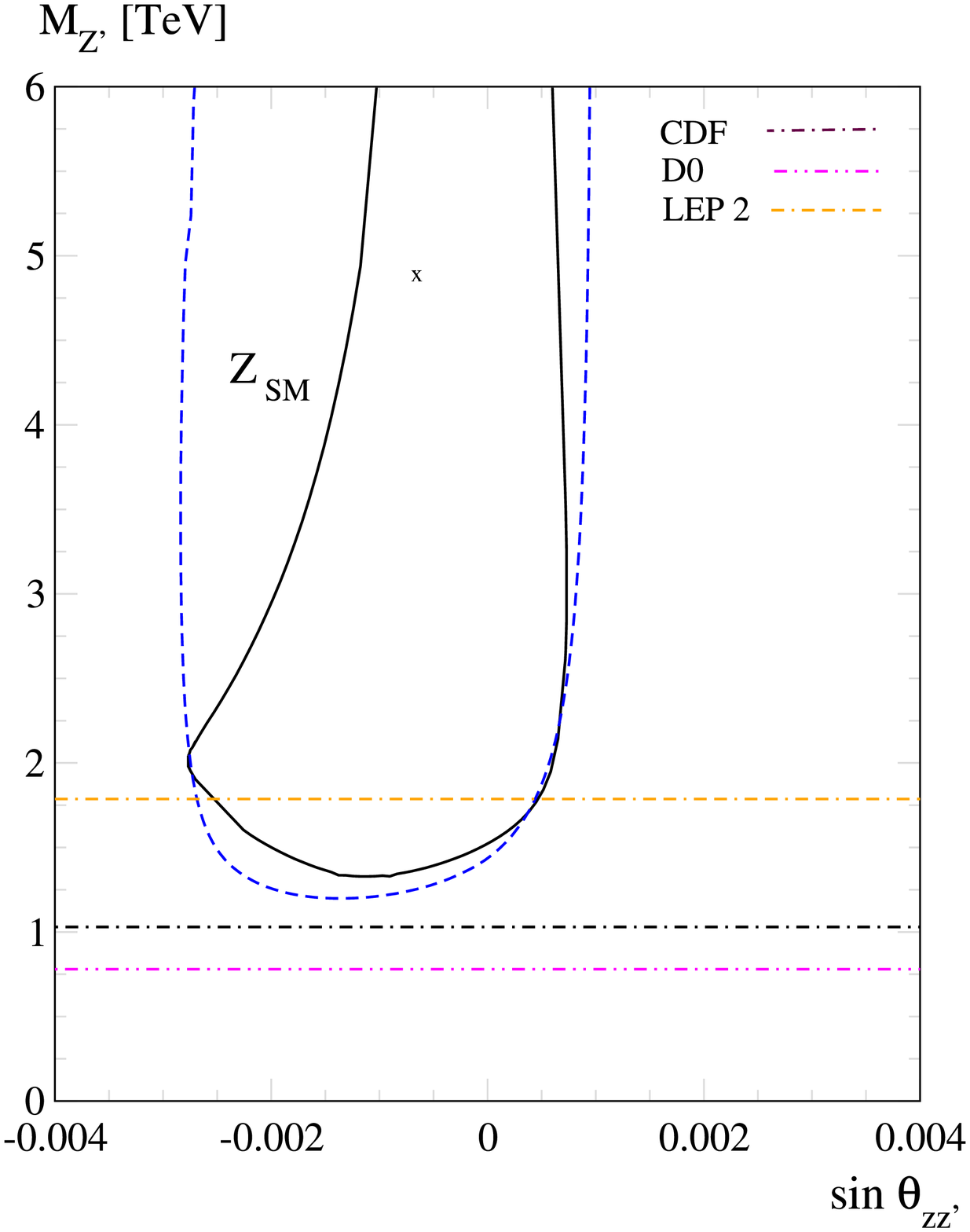} & \hspace{55pt} \includegraphics[scale=0.33]{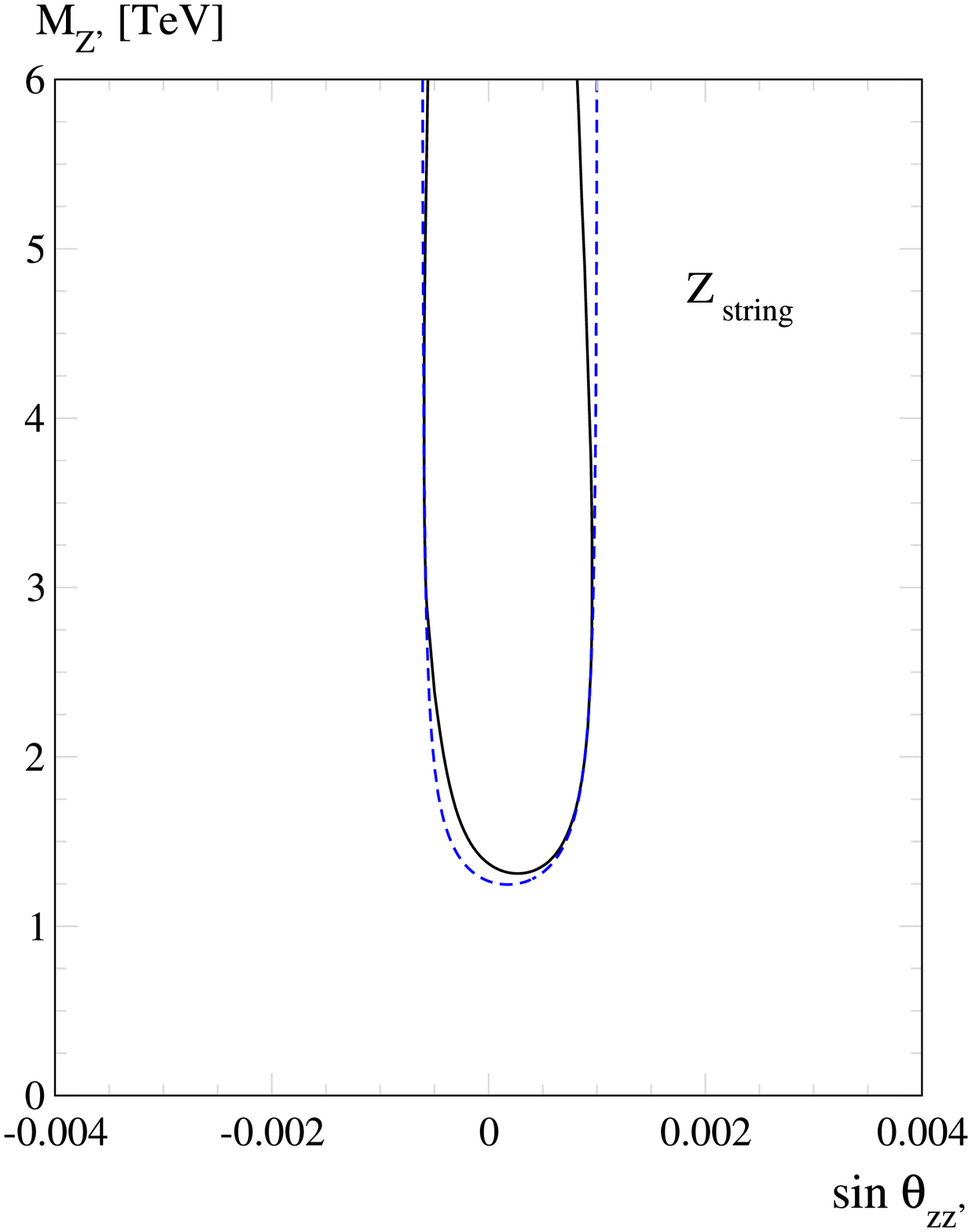} \vspace{10pt} \\
\end{tabular}
\caption{90\% C.L. contours in $M_{Z'}$ vs. $\sin \theta_{ZZ'}$ for
  various models. See the text for details.}
\label{Contours2}
\end{figure}

Figures~\ref{Contours1} and \ref{Contours2} show 90\% C.L. exclusion contours for all
models. The solid lines specify use of the constraint $\rho_0 = 1$
while the dashed lines are for $\rho_0$ free. We also show the extra
constraints for the specific Higgs sectors discussed above. These are
represented by the dotted and long-dashed lines. The numbers in the
plots refer to the values of $\tau$ or $\omega$, from ranges given in
eq.~\ref{Cvalues}. The best fit locations are indicated by an {\tt "x"}. The lower limits from CDF (dot-dashed), D\O\ (double-dot-dashed) and LEP~2 (dot-double-dashed) given in Table~\ref{Limits} are also shown.

\section*{Acknowledgments} 
The work at IF-UNAM is supported by CONACyT project 82291--F. The work of P.L.~is supported by the IBM Einstein Fellowship and by NSF grant PHY--0503584.

\end{document}